\documentclass[a4paper,notitlepage]{article}

\usepackage[dvips]{graphicx}

\newcommand{\be}{\begin{equation}}

\newcommand{\ee}{\end{equation}}

\title{Distribution of the largest eigenvalues of the L\'{e}vi -
Smirnov ensemble.} 
\author{Waldemar Wieczorek\\ \\ M. Smoluchowski Institute of Physics\\                                          
Jagellonian University\\                                                        
Reymonta 4, 30-059 Cracow, Poland}

\begin{document}

 \maketitle
\begin{abstract}
We calculate the distribution of the $k$-th largest eigenvalue in
the  random matrix L\'{e}vi-Smirnov (LSE) ensemble, 
using the spectral dualism between LSE and chiral Gaussian Unitary Ensemble
(GUE). Then we  reconstruct universal spectral  oscillations and 
we investigate an asymptotic behavior of the spectral distribution.
\end{abstract}

\section{Introduction}

Random matrix ensembles provide a powerful and generic formalism allowing 
addressing several statistical problems  of the energy spectra in various complex quantum systems. In most applications, the ensembles used belong 
to the basin of attraction of the Gaussian ensembles.
Relatively few is known about the mathematical properties of the 
matrix ensembles with so strongly fluctuating elements, that the average 
and the variance of the ensemble diverge. 
Such ensembles can be viewed as the matrix generalization of the stable 
probability distributions of the probability theory, usually known 
as L\'{e}vy distributions.    
Historically, such ensembles of matrices were proposed by Bouchaud and Cizeau
~\cite{BOU-CIZ}. Matrix elements of the corresponding matrices were 
drawn from stable, one-dimensional probability distributions. 
  This construction, however, was breaking the rotational invariance of the  
 ensemble, imposed in most Gaussian-like and polynomial matrix models. 
Recently, borrowing from the mathematical concepts of free random variables 
analysis~\cite{VOICULESCU}, a new class of stable, rotationally invariant 
matrix models was introduced in ~\cite{man5}. 
Exploiting the Coulomb gas analogy, the authors~\cite{man5}
 propose a general method 
of constructing explicit matrix ensembles characterized by stable 
power spectra, with asymptotic behavior $\lambda^{-1-\alpha}$.
 The stability condition means, that the matrix convolution 
of the identical  independent ensembles with power-like spectra 
exhibits, modulo the rescaling and shift, the same power-like 
spectral distribution. In this sense, the ensembles  provide a generalization
of central limit theorem of Gaussian ensembles, where 
e.g. the convolution of two independent identical Gaussian Unitary Ensembles
(GUE)  with unit variance is also the Gaussian Unitary Ensemble, 
however with the semi-circle law rescaled by $\sqrt{2}$ 
in the longitudinal direction.  
Due to the growing interest 
in stochastic processes with long (sometimes called also  heavy or fat)
tails, intermittency and anomalous diffusion, the theory of stable 
random matrices may form an  important role in 
describing generic features of complex systems using new 
concepts of universality.
Very recently, new multi-critical random ensembles labeled by 
{\em continuous} scaling exponents  were proposed~\cite{JANIK}, with 
eigenvalue density near zero behaving like $|x|^{a}$. 
Similar continuous series of universality class 
for {\large} eigenvalues 
appear 
naturally in stable random matrix ensembles with a non-compact support~\cite{man5}. 
It is  plausible  to   conjecture strongly~\cite{JANIK}, 
that by  simple mapping  
the scaling behavior of small eigenvalues 
of the continuous multi-fractal regime  can be related 
to the large eigenvalue behavior of the L\'{e}vy random matrix models.

In this work, we study  one of the simplest ensembles of L\'{e}vy matrices, 
so-called L\'{e}vy-Smirnov matrix model. The name origins from the 
fact, that the model constitutes a matrix analog of the 
L\'{e}vy-Smirnov probability distribution (\cite{FELLER}),
i.e. the spectrum for large eigenvalues is  power-like with $\alpha=1/2$.
In the next chapter, we remind the definition and some properties 
of the LS ensemble.  
Then,  we  exploit the duality of this model~\cite{man5} to
 the massless chiral Gaussian Unitary Ensemble~\cite{CHIRAL}.
Since the chiral GUE plays a crucial role in describing the universal 
properties of low-lying eigenvalues of QCD Dirac operator, 
an impressive number of analytical tools and methods was developed 
in few last years. Using the aforementioned duality and some of these
 methods~\cite{dag}, we  calculate the distribution  
of $k$-th largest eigenvalue in the L\'{e}vy-Smirnov  ensemble  (hereafter LSE),
and show explicit results for $k=1,2,3,4$. Finally, 
 we analyze an asymptotic behavior of the spectral function,
demonstrating  how this behavior can be obtained from the 
microscopic spectral function. We close the paper with the discussion.

\section{L\'{e}vi-Smirnov ensemble.}
The Random Matrix Theory is usually  defined by the partition
 function
\be 
Z = \int dM e^{-N {\rm Tr} V(M)} \, , 
\ee
where $M$ is $N$ by $N$ matrix and $V(M)$ is a potential, 
which, in general,  can be even  a  non-analytical function of $M$.
Spectral distribution of the ensemble is defined as 
\be
\rho(\lambda)=\frac{1}{N}<{\rm Tr } \delta (\lambda -M)>
\ee
where the averaging is done with the measure defined by the partition 
function. The Gaussian Unitary Ensemble (GUE) is characterized 
by quadratic potential $V(M)=1/2M^2$ (we put variance to 1),
 leading to Wigner semi-circle law
for the eigenvalues, $\rho(\lambda)=1/{2\pi}\sqrt{4-\lambda^2}$, 
i.e. the eigenvalues are localized on the compact interval $[-2,2]$. 
    
L\'{e}vy-Smirnov ensemble is characterized by a rather non-trivial
 potential ~\cite{man5}
\be
V(M)= \exp [-N {\rm Tr}( M^{-1} +\ln M)]
  \ee
leading to the spectral function 
\be
\rho_{LS}(\lambda)=\frac{1}{2\pi}\frac{\sqrt{4\lambda-1}}{\lambda^2}
\label{rhols}
\ee
localized on the non-compact support $[1/4, \infty]$, 
with asymptotic behavior for large spectra $1/\lambda^{1+\alpha}$, with 
$\alpha=1/2$.  This behavior explains the name of the ensemble, since 
the probability distribution function for stable density with similar 
$\alpha$ and similar maximal asymmetry bears this name
\be
p_{LS}(x)=\frac{1}{\sqrt{2\pi}}\exp(-1/(2x)) x^{-3/2} \,\,\,\,\,\,\,x>0 
\label{LSprob}
\ee
Note that LS matrix ensemble is {\em not} obtained simply by populating the matrix elements from the probability distribution (\ref{LSprob}).  
 
\section{A duality between the Chiral and LS ensembles.} 

 The LS  ensemble  is one of the simplest example of 
L\'{e}vi matrices, because there exist  an  exact, analytic 
expression for $V(M)$.
 This model has an unitary symmetry, so we can diagonalize $M \rightarrow U\Lambda U^{\dagger}$ and, after  integration 
 over the angular variables, rephrase  the 
 partition in terms of the eigenvalues only, 
the last term being the Vandermonde determinant 
coming from the Jacobian ( see e.g. \cite{MEHTA} for standard discussion)
\be
Z \equiv\int^{\infty}_{0}\prod_{i}d\lambda_{i}\left(
\frac{e^{-N/\lambda_{i}}}{\lambda_{i}^{2N}}\right)\prod_{i>j}\left(\lambda_{i}-\lambda_{j}\right)^{2} \, .
\ee 

In this section we show that the distribution of eigenvalues in
LSE behaves, up to trivial transformation, in
the same way as  a distribution of the eigenvalues for
the  chiral  Gaussian Unitary ensemble. We will
show, that the equations describing both LS  and Gaussian ensembles
can be proved by the same method, using almost identical expressions.

  We assume, that $N$ eigenvalues are  ordered
$(\lambda_{1}>\lambda_{2}>\ldots>\lambda_{N})$.  An  unnormalized
joint probability density $\Omega_{N,k}$ defined as in \cite{dag} in
our case takes a form: 
\be \label{wz1}
\Omega_{N,k}(\lambda_{1},\lambda_{2},\ldots,\lambda_{k}) = {\cal
N} \frac{1}{(N-k)!}\int^{\lambda_{k}}_{0} d\lambda_{k+1} \ldots
\int^{\lambda_{k}}_{0} d\lambda_{N} \prod_{i=1}^{N}\left(
\frac{e^{-N/\lambda_{i}}}{\lambda_{i}^{2N}}\right)\prod_{i>j}\left(\lambda_{i}-\lambda_{j}\right)^{2} \, ,
\ee
 where $\cal N$ stands for the normalization:
 \be {\cal N} =
\frac{1}{N!}\int^{\infty}_{0}\prod_{i=1}^{N}d\lambda_{i}\left(
\frac{e^{-N/\lambda_{i}}}{\lambda_{i}^{2N}}\right)\prod_{i>j}\left(\lambda_{i}-\lambda_{j}\right)^{2}.
\ee
The main difference between eq. (\ref{wz1}) and its analog in chGUE 
lies  in the range of integration. In LS
 case we are interested by the large eigenvalues so we integrate over an area with eigenvalues {\em smaller} than $\lambda_{k}$.

 Using methods originally developed  for chGUE in \cite{dag},
 we can calculate 
the unnormalized joint probability taking a 
microscopic limit. To do this,  we change variables in the integral: $\lambda_{i}=1/x_{i}, \mbox{ for } i
= k+1, \ldots, N$ and we shift variables  $x_{i}\rightarrow x_{i}+1/\lambda_{k}$. This  leads us to the expression: 

\begin{eqnarray}
 \Omega_{N}(\{\lambda_{k}\}) &=& \frac{{\cal N}}{(N-k)!}
e^{-\frac{N(N-k)}{\lambda_{k}}}
\prod_{i=1}^{k}\frac{e^{-N/\lambda_{i}}}{\lambda_{i}^{2N}}
\prod_{i<j}^{k}\left(\lambda_{i}-\lambda_{j}\right)^{2} \cdot \\
\nonumber&&\cdot \int^{\infty}_{0} \prod_{i=k+1}^{N}\, dx_{i}
e^{-N x_{i}} \lambda_{k}^{2}x^{2}_{i}
\prod_{j=1}^{k-1}(1-\lambda_{j}x_{i}-\lambda_{j}/\lambda_{k})^{2}
\prod_{i>j\geq k+1}\left(x_{i}-x_{j}\right)^{2}.
\end{eqnarray}

Now we  take a microscopic limit. 
Motivated by \cite{dag,man5,JANIK}
 we go with both $\lambda_{i}$ and $N$ to infinity  keeping
$\xi^{2}_{i}=\lambda_{i}/4N^{2}$ fixed. It is useful to change notation to $\Lambda_{i} = 1/\xi_{i}$, which help us to find the analogy mentioned before.  We calculate  rescaled function $\Omega_{k}$ called (according to \cite{dag}) $\omega_{k}(\Lambda_{1},\ldots,\Lambda_{k})$:
\begin{eqnarray}
\omega_{k}(\Lambda_{1},\ldots,\Lambda_{k})&=&\lim_{N\rightarrow\infty}\left( \prod_{i=1}^{k}\frac{4N^{2}}{\Lambda^{3}_{i}}\right)
\Omega_{N,k}\left(\frac{4N^{2}}{\Lambda_{1}^{2}},\frac{4N^{2}}{\Lambda_{2}^{2}},\ldots,\frac{4N^{2}}{\Lambda_{k}^{2}}\right)=\\
&=&\lim_{N\rightarrow\infty}\left(\prod_{i=1}^{k}\frac{4N^{2}}{\Lambda^{3}_{i}} \right)
{\cal N}\frac{e^{-\Lambda^{2}_{k}/4}}{(N-k)!} \prod_{i=1}^{k}
(\Lambda_{i}^{2})^{2}
\prod_{i<j}^{k-1}\left(\Lambda^{2}_{i}-\Lambda^{2}_{j}\right)^{2}
\cdot (4N^{2})^{-2k^{2} + k(k-1)} \times\\
\nonumber && \times \int^{\infty}_{0} \prod_{i=k+1}^{N}\, dx_{i}
e^{-N x_{i}} x^{2}_{i}
\prod_{j=1}^{k-1}\left(x_{i}-\frac{\mu_{j}^{2}}{4N^{2}}\right)^{2}
\prod^{N}_{i>j\geq k+1}\left(x_{i}-x_{j}\right)^{2},
\end{eqnarray}
where we have introduced $\mu_{j}^{2} =
\Lambda^{2}_{j}-\Lambda^{2}_{k}$.

Now, we are able to write this expression using the partition functions for
 chGUE $\cal Z$ in the microscopic limit:
 \be
\label{wz2}
\omega_{k}(\Lambda_{1},\ldots,\Lambda_{k}) = \mbox{const }
e^{-\Lambda^{2}_{k}/4} \prod_{i=1}^{k} \Lambda_{i}
\prod_{i<j}^{k-1}\left(\Lambda^{2}_{i}-\Lambda^{2}_{j}\right)^{2}
\frac{{\cal Z}_{2}(\mu_{1}, \mu_{1},\mu_{2}, \mu_{2}, \ldots ,
\mu_{k-1}, \mu_{k-1})}{{\cal Z}_{0}(0)} \, .
 \ee
This expression shows the 
deep analogy between  chGUE and LS  ensembles.
With the help of the above expression, in the next section, we will
derive   
the rescaled joint 
 probability distribution of the $k$ largest eigenvalues in the LSE.

 \section{Distribution of the $k$-th eigenvalue}

In this section we derive  an analytical
 expression for the distribution of $k$-th eigenvalue.
This distribution is defined as 
\be p_{k}(\Lambda)
=
\int_{0}^{\Lambda}d\Lambda_{1}\int^{\Lambda}_{\Lambda_{1}}d\Lambda_{2}
\ldots \int^{\Lambda}_{\Lambda_{k-2}}d\Lambda_{k-1}\;
\omega_{k}(\Lambda_{1},\ldots,\Lambda_{k-1},\Lambda)\, . \ee

Using (\ref{wz2}) and  changing variables
$x_{i}=\sqrt{\Lambda_{i}^{2}-\Lambda_{k}^{2}}$ for $i=1,\ldots
 k-1$ we can write the distribution in the form: \be
p_{k}(\Lambda)= \mbox{ const }
\int_{0}^{\Lambda}dx_{1}\int^{x_{1}}_{0}dx_{2} \ldots
\int^{x_{k-2}}_{0}dx_{k-1}\;e^{-\Lambda^{2}/4} \prod_{i=1}^{k}
x_{i} \prod_{i<j}^{k-1}\left(x^{2}_{i}-x^{2}_{j}\right)^{2} {\cal
Z}_{2}(x_{1}, x_{1}, \ldots , x_{k-1}, x_{k-1}). \ee
 
An expression showed above contains the partition function for massive
chiral GUE model, with twofold degeneracy for masses.
 With the help of the formula for this partition function, 
 proven in  \cite{akemann}, we can rewrite our final result.
In our variables, equation (2.1) from \cite{akemann} reads
  \be
\prod_{i<j}^{k-1} \left( (x^{2}_{i}-x^{2}_{j}\right)^{2} {\cal
Z}_{2}^{(2k)}(x_{1}, x_{1}, \ldots , x_{k-1}, x_{k-1})= \mbox{
const } \det_{1\leq a,b\leq k} \left[{\cal Z}_{2}^{(2)}
(x_{a},x_{b})\right] \; .
\label{ake}
 \ee

Using (\ref{ake})  we 
write down our final result:
 \be p_{k}(\Lambda)= \mbox{ const }
\int_{0}^{\Lambda}dx_{1}\int^{x_{1}}_{0}dx_{2} \ldots
\int^{x_{k-2}}_{0}dx_{k-1}\;e^{-\Lambda^{2}/4} \prod_{i=1}^{k}
x_{i} \det_{1\leq a,b\leq k} \left[{\cal Z}_{2}^{(2)}
(x_{a},x_{b})\right] \;. 
\ee 
where the  partition function for two masses
was already calculated in  \cite{akemann}):
 \be {\cal Z}_{2}^{(2)}
(x_{a},x_{b}) = \frac{x_{b} I_{2}(x_{a}) I_{3}(x_{b}) - x_{a} I_{2}(x_{b})
I_{3}(x_{a})}{(-x_{a}^{2}+x_{b}^{2})} \, . \ee

Now, we can go back to the
original rescaled variable $\xi$ (proportional to un-scaled
$\lambda$). Using substitution $\Lambda=\frac{2}{\sqrt\xi}$ we can
write the expression for the probability density of the $k$-th
eigenvalue: 
\be
 \label{result}
p_{k}(\xi)=\frac{\mbox{const}}{\xi^{2}}
\int_{0}^{2/\sqrt{\xi}}dx_{1}\int^{x_{1}}_{0}dx_{2} \ldots
\int^{x_{k-2}}_{0}dx_{k-1}\;e^{-1/\xi} \prod_{i=1}^{k-1} x_{i}
\det_{1\leq a,b\leq k} \left[{\cal Z}_{2}^{(2)}
(x_{a},x_{b})\right] \;.\ee

 This probability needs to be
normalized. We  calculate this expression for several cases
$k=1,2,3,4$. The case $k=1$ is very simple:
\be
p_{1} (\xi) = \frac{1}{\xi^{2}} e^{-1/\xi} \, .
\ee
 For higher $k$'s
(greater than 2) calculations are more complicated and cannot be
performed by   analytical methods, so we use the numerical method instead.

Figure 1 shows the result for probability distribution for  
first four largest eigenvalues of the LS ensemble.

\begin{figure}[h!]
\begin{center}
\includegraphics[width=6.5cm]{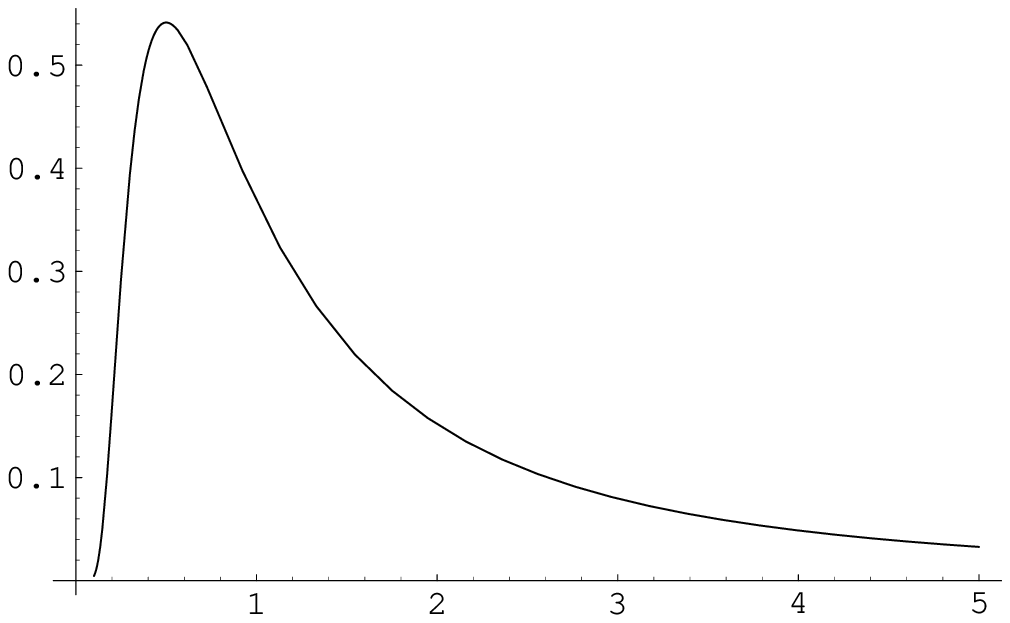}
\includegraphics[width=6.5cm]{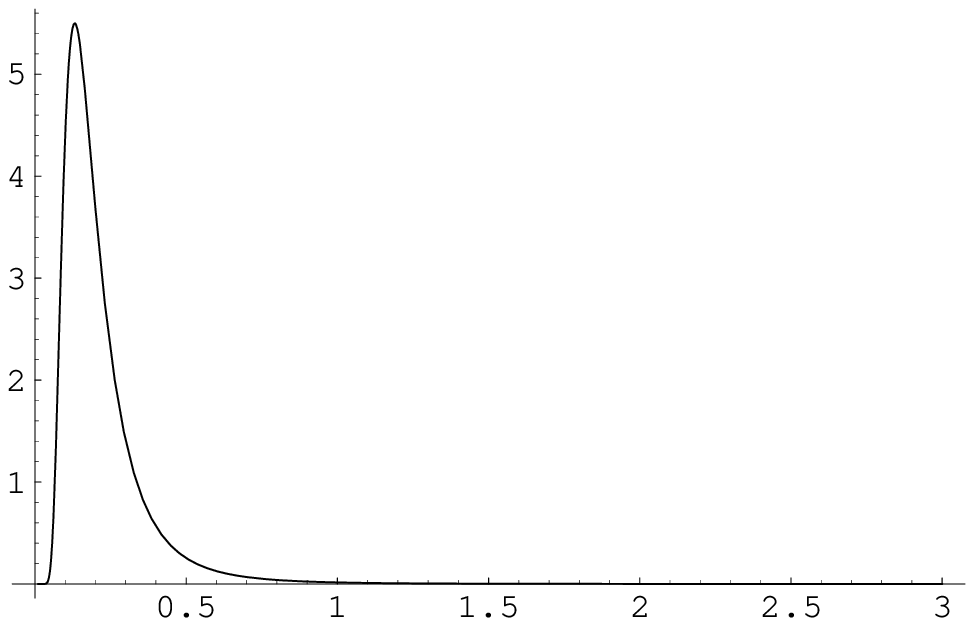}
\includegraphics[width=6.5cm]{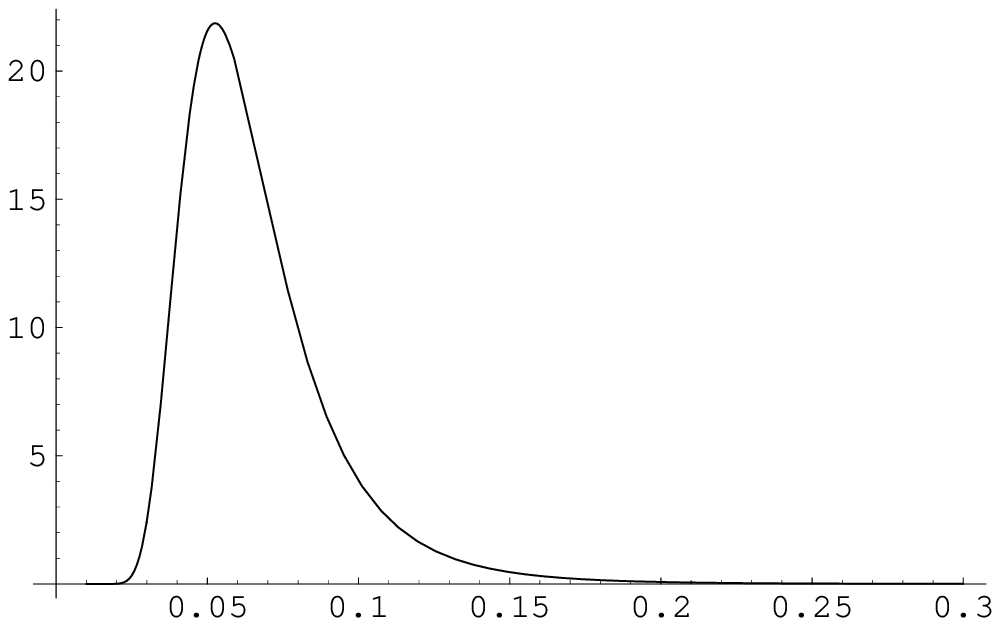}
\includegraphics[width=6.5cm]{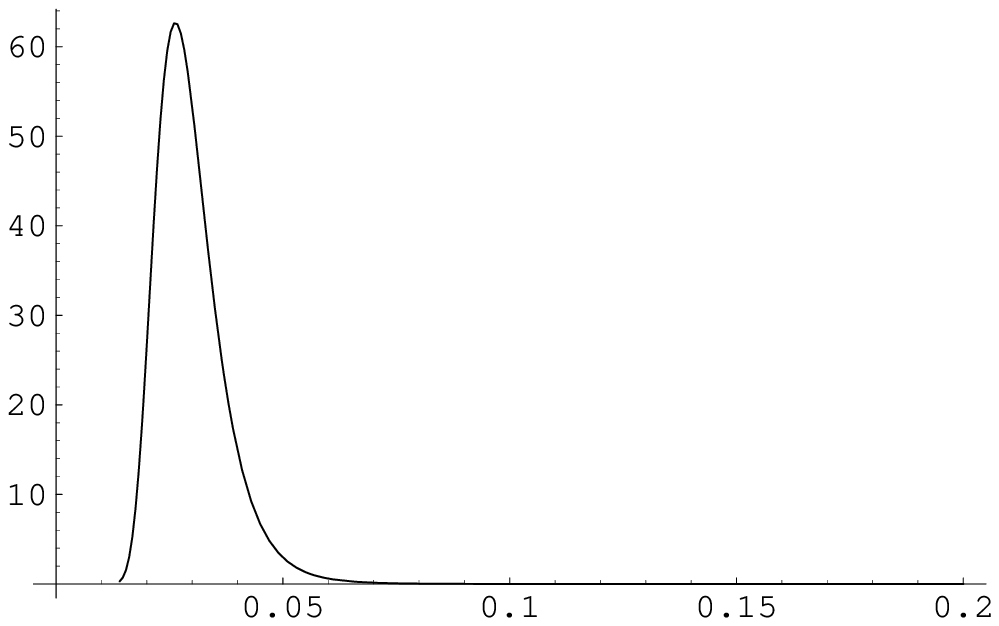}
\caption{The normalized
distribution of the first four eigenvalues $k=1,2,3,4$ (from left to
right).}
\end{center}
\end{figure}

By definition, 
the  sum of $p_k$ distributions 
reproduces  the microscopic spectral density for LSE:
 \be
\label{wz} \rho_{S} (\xi) = \sum_{k=1}^{\infty} p_{k} (\xi)
= (\xi)^{-2} \left( J_{0}^{2}(2/\sqrt{\xi}) +
J_{1}^{2}(2/\sqrt{\xi})\right)\, .\ee 
calculated in~\cite{man5}. 
Note also, that the change of variables rewrites this density 
in terms of microscopic spectral density 
of chiral GUE, calculated by \cite{VER_ZAH}.

As an independent check of our formulae, 
we  can approximately reconstruct $\rho
(\xi)$ by adding first four distributions. Because the $\rho (\xi)$
asymptotic behavior is like $1/\xi^{2}$,  it is more convenient  to plot
$\rho \cdot \xi^{2}$ instead $\rho$. Figure~2  shows how 
 oscillatory universal pattern for large eigenvalues is reproduced 
by the sum of the four largest eigenvalues. 

\begin{figure}[h]
\label{wykr}
\begin{center}
\includegraphics[width=12cm]{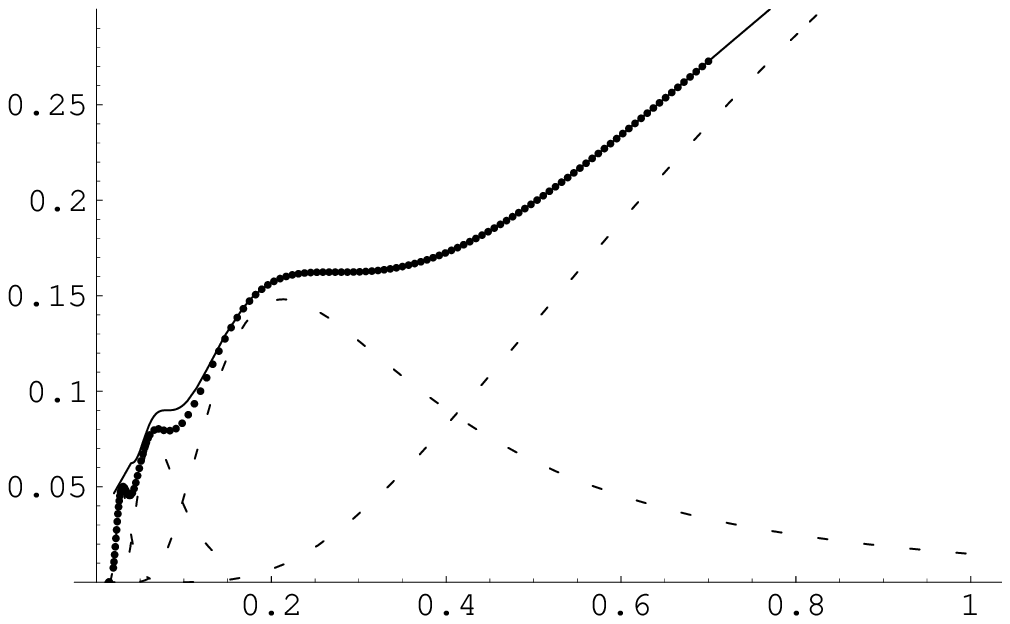}
\caption{The distribution of the first four eigenvalues $p_{k}(\xi)$ for $k=1,2,3,4$ (dashed lines)
 their sum (dotted line)
 and $\rho_{S}(\xi)$ (solid line) - all function are multiplied by $\xi^{2}$.}
\end{center}
\end{figure}

\section{Asymptotic behavior}

In this section, for completeness, 
we reconstruct the asymptotic behavior 
of the distribution function $\rho(\lambda)$ for LS ensemble,  
given by (\ref{rhols}). 
The asymptotic behavior
in infinity is  $\lambda^{-3/2}$:
 \be
\rho_{LS}(\lambda)=\frac{1}{2\pi} \frac{\sqrt{4\lambda
-1}}{\lambda^{2}} \approx \frac{1}{\pi} \frac{1}{\lambda^{3/2}} \,
. \ee 
We can reconstruct this result from equation (\ref{wz}),
which we got after taking a microscopic limit. This means that we
blow up the region  close to infinity,  because we took the limit
$\lambda\rightarrow \infty$ keeping $\frac{\lambda}{N^{2}}$ fixed.
This corresponds to  asymptotic behavior of the scaling functions
 for $\xi$ close to 0. We
use well-known property  of the Bessel functions (for $x$ real)
 \be J_{\nu}(x)
 \approx \sqrt{\frac{2}{\pi x}} \left( \cos (x - \frac{1}{2} \nu \pi -
 \frac{1}{4}) + {\cal O}(x^{-1}) \right)
\mbox{, for } x\approx 0\, .\ee

Applying this result to the (\ref{wz}) we obtain the expected result:
\be
\rho_{S} (\xi) \approx \frac{1}{\xi^2} \frac{\sqrt{\xi}}{\pi} \left( \cos^2 (
\frac{2}{\sqrt{\xi}} - \frac{\pi}{4}) + \sin^2  (
\frac{2}{\sqrt{\xi}} - \frac{\pi}{4}\right) = \frac{1}{\pi} \frac{1}{\xi^{3/2}}\, .\ee

\section{Discussion}

The main result of this paper is the equation (\ref{result}), which allows 
the calculations of  $p_{k}(\xi)$ in an explicit way.
 The
 analytical form of these probability distributions can be used for comparison to 
 similar empirical probability distributions of the {\it largest} 
 eigenvalues of stable matrix ensembles, alike it is done in QCD lattice analysis
 for the {\it smallest} eigenvalues \cite{WETTIG}.
 To obtain this result we used well-known methods  in chiral random 
matrix models \cite{dag,akemann} and we exploited a non-trivial  duality 
between the chiral Gaussian ensemble and LS ensemble.
Naively, both ensembles are very different in nature. 
Chiral Gaussian ensemble is governed by a  simple quadratic potential and  
exhibits a  non-trivial symmetry due to the off-diagonal block structure.
Its spectrum is localized on the compact support, with novel (comparing to 
non-chiral Gaussian ensemble) microscopic oscillations in the vicinity of zero.
LS ensemble belongs to non-Gaussian stability region, potential is a 
non-analytic function of $M$, and spectrum has an infinite support.
On the basis of the duality known between $\alpha$ and $1/\alpha$ 
distributions for stable one-dimensional probability distributions~\cite{ZOLO} 
one may expect  the similar duality between the bulk 
spectral function for GUE and LSE. Indeed, the semi-circle 
law transmutes into (\ref{rhols}), by  
changing    $\lambda^2$ into $1/\lambda$.  This transformation misses
however the chiral character of the dual ensemble. It is the equivalence 
of the partition functions at the microscopic level, that allows to transmute
all non-trivial spectral properties of the chiral ensemble 
in the vicinity of zero to the spectral, universal properties of LS ensemble   
for large eigenvalues. 
In some sense, LS model is unique, allowing, despite having divergent
mean and variance, to use the methods reserved for finite variance and 
compact support models. In general, the study of microscopic 
properties of L\'{e}vy  ensembles requires mapping described 
in \cite{JANIK}  and probably gives only approximate universal scalings.
It is interesting to study  other spectral statistics
of the L\'{e}vy ensembles, especially in connection  with the critical 
level statistics
in disordered media~\cite{KRA_MUT}.   

\section*{Acknowledgments:}
This work was supported by the Polish Government Projects (KBN) 2P03B 09622
(2002-2004).

\end{document}